\documentclass[11pt,leqno,titlepage]{article}

\makeatletter%
\def\nottoobig#1{{\hbox{$\left#1\vcenter to1.111\ht\strutbox{}\right.\n@space$}}}
\makeatother%

\def\union{\cup}
\def\inter{\cap}

\newlength{\filength}
\newsavebox{\gcbox}
\sbox{\gcbox}{\framebox[\filength]{\rule{0ex}{2ex}}}

 \newtheorem{theorem}{Theorem}[section]
 \newtheorem{corollary}[theorem]{Corollary}

\newcommand{\qedblob}{\mbox{\rule[-1.5pt]{5pt}{10.5pt}}}
\def\literalqed{{\ \nolinebreak\hfill\mbox{\qedblob\quad}}}

\def\qed{\literalqed}

 \newtheorem{proposition}[theorem]{Proposition}

 \newtheorem{definition}[theorem]{Definition}

\newcommand{\naturalnumber}{{\rm I\!N}}

\newcommand\seq{\subseteq}
\newcommand{\sharpp}{{\rm \#P}}

\newcommand{\sat}{{\rm SAT}}

\newcommand{\parityp}{{\rm \oplus P}}
\newcommand{\up}{{\rm UP}}

\newcommand{\fewp}{{\rm FewP}}
\newcommand{\coup}{{\rm coUP}}

\newcommand{\p}{{\rm P}}

\newcommand{\ga}{{\rm GA}}
\newcommand{\coam}{{\rm coAM}}
\newcommand{\bne}{{\rm BNE}}
\newcommand{\np}{{\rm NP}}
\newcommand{\rc}{{\rm RC}}

\newcommand{\pp}{{\rm PP}}

\newcommand{\listp}{{\rm ListP}}

\newcommand{\poly}{{{\rm poly}}}

\newcommand{\few}{{{\rm Few}}}

\newcommand{\implies}{\, \Longrightarrow \ }

\newcommand{\sigmastar}{{\Sigma^\ast}}

\newcommand{\bigo}{{\protect\cal O}}

\newcommand{\condition}{\,\nottoobig{|}\:}
 
\def\land{{\; \wedge \;}}

\title{
Restrictive Acceptance Suffices for Equivalence
Problems\footnote{Revises Friedrich-Schiller-Universit\"at Jena
Technical Report Math/Inf/96/13.}}

\author{
{Bernd Borchert}\thanks{
Email: {\tt bb@math.uni-heidelberg.de}.
}
\\Mathematisches Institut
\\Universit\"at Heidelberg
\\69120 Heidelberg, Germany
\and
{Lane A. Hemaspaandra}\thanks{
Email: {\tt lane@cs.rochester.edu}.
Supported in part by 
grants NSF-CCR-9322513,
NSF-INT-9513368/DAAD-315-PRO-fo-ab,
and
NSF-INT-9815095/DAAD-315-PPP-g\"{u}-ab.
Work done in part while visiting 
Friedrich-Schiller-Universit\"at Jena.
}
\\Department of Computer Science
\\University of Rochester
\\Rochester, NY 14627, USA 
\and
{J\"org Rothe}\thanks{
Email: {\tt rothe@informatik.uni-jena.de}.
Supported
in part by grants
NSF-INT-9513368/DAAD-315-PRO-fo-ab and
NSF-INT-9815095/DAAD-315-PPP-g\"{u}-ab,
and
a 
NATO Postdoctoral Science Fellowship
from the Deut\-scher Aka\-de\-mi\-scher Aus\-tausch\-dienst
(``Ge\-mein\-sames Hoch\-schul\-sonder\-pro\-gramm~III 
von Bund und L\"andern'').
Work done in part while visiting the University of Rochester and 
Le Moyne College.
}
\\Institut f\"ur Informatik
\\Friedrich-Schiller-Universit\"at Jena
\\07740 Jena, Germany
} %

\date{}

\newcommand{\ep}{{\rm EP}}
\newcommand{\es}{{\rm ES}}

\newcommand{\npnp}{{\np^{\rm NP}}}

\newcommand{\ceqp}{{{\rm C_{\!=\!}P }}}
\newcommand{\ceqphalf}{{{\rm C_{\!= =\!}P[half] }}}
\newcommand{\spp}{{{\rm SPP}}}

\newcommand{\fp}{{{\rm FP}}}

\newcommand{\psharpp}{{\p^\sharpp}}

\begin{document}

\typeout{WARNING:  BADNESS used to supress reporting.  Beware.}
\hbadness=3000%
\vbadness=10000 %

\bibliographystyle{plain}

{%
\maketitle}

\begin{abstract}
{One way of suggesting that an NP problem may not be NP-complete is to
show that it is in the class UP\@.  We suggest an analogous new
approach---weaker in strength of evidence but more broadly
applicable---to suggesting that concrete~NP problems are not
NP-complete.  In particular we introduce the class EP, the subclass of
NP consisting of those languages accepted by NP machines that when
they accept always have a number of accepting paths that is a power of
two.  Since if any NP-complete set is in EP then all NP sets are in
EP, it follows---with whatever
degree of strength one believes that EP differs from NP---that 
membership in EP can be viewed 
as evidence
that a problem is not NP-complete.

We show that the negation equivalence 
problem for OBDDs (ordered binary 
decision 
diagrams~\cite{for-hop-sch:c:equivalence,bry:j:obdd-survey}) 
and the interchange
equivalence problem for 2-dags are in EP\@.  We also
show that 
for boolean 
negation~\cite{har:b:boolean-survey}
the equivalence problem 
is in $\ep^{\np}$, thus tightening the existing $\np^\np$
upper bound.  We show that 
FewP~\cite{all-rub:j:print},
bounded ambiguity polynomial time, is contained in EP, a result that 
is not known to follow from the previous SPP upper bound.  
For the three problems and classes just mentioned with
regard to EP, no proof of 
membership/containment in UP is known,
and for the problem just mentioned with
regard to EP${}^\np$, no proof of 
membership in UP${}^\np$ is known.  Thus, EP is indeed
a tool that gives evidence against 
NP-completeness in natural cases where UP cannot currently be applied.

} %
\end{abstract}
\let\BLS=\baselinestretch

\makeatletter
\newcommand{\niceonespacing}{\let\CS=\@currsize\renewcommand{\baselinestretch}{1.1}\tiny\CS}\newcommand{\nicetwospacing}{\let\CS=\@currsize\renewcommand{\baselinestretch}{1.2}\tiny\CS}
\newcommand{\nicethreespacing}{\let\CS=\@currsize\renewcommand{\baselinestretch}{1.3}\tiny\CS}
\newcommand{\singlespacingplusplus}{\let\CS=\@currsize\renewcommand{\baselinestretch}{1.35}\tiny\CS}
\newcommand{\nicefourspacing}{\let\CS=\@currsize\renewcommand{\baselinestretch}{1.4}\tiny\CS}
\newcommand{\nicefivespacing}{\let\CS=\@currsize\renewcommand{\baselinestretch}{1.5}\tiny\CS}
\newcommand{\nicesixspacing}{\let\CS=\@currsize\renewcommand{\baselinestretch}{1.6}\tiny\CS}
\makeatother

\pagestyle{plain}
\sloppy

\sloppy

\setcounter{footnote}{0}

\section{Introduction}
NP languages can be defined via machines that reject by having 
zero accepting
paths, and that accept by having their number of accepting paths
belong to the set $\{1,2,3,\ldots\}$.  A number of researchers have 
sought to refine the class NP by shrinking the 
path-cardinality set signifying acceptance, while retaining the requirement
that rejection be associated with having zero accepting paths.  
We will call any such class a {\em restricted counting
class}.  The most common restricted counting classes in the
literature are 
random polynomial time (usually denoted R or RP) 
and ambiguity-bounded classes such as UP and FewP\@.  
Ambiguity-bounded classes will be of central interest
to us in the present paper.

Valiant's
class UP (unambiguous polynomial 
time)~\cite{val:j:checking}, which is known to differ from 
P exactly if one-way functions 
exist~\cite{gro-sel:j:complexity-measures},
has the acceptance set $\{1\}$, and so is a 
restricted counting class.   Acceptance sets of 
the forms $\{1,2,3,\ldots,n^{\bigo(1)}\}$ and
$\{1,2\}$, 
$\{1,2,3\}$, $\ldots$ define, respectively,
the class FewP~\cite{all-rub:j:print}
and the classes
$\up_{\leq 2}$, 
$\up_{\leq 3}$, $\ldots$~\cite{bei:c:up1}, and thus 
these too are restricted counting classes.
(Note: 
\mbox{$\up \seq \up_{\leq 2} \seq 
\up_{\leq 3} \seq \cdots \seq \up_{\bigo (1)} \seq 
\fewp \seq \np$}, where 
$\up_{\bigo (1)} = \bigcup_{k \geq 1} \up_{\leq k}$.)~~These 
classes are also connected to the existence of one-way functions
and have been extensively studied in a wide variety of 
contexts, such as 
class containments~\cite{fen-for-kur:j:gap,koe-sch-tod-tor:j:few}, 
complete sets~\cite{hem-jai-ver:j:up-turing},
reducibilities~\cite{hem-hem:j:quasi}, 
boolean hierarchy equivalences~\cite{hem-rot:j:boolean},
complexity-theoretic analogs of Rice's 
Theorem~\cite{bor-ste:c:rice},
and 
upward 
separations~\cite{rao-rot-wat:jCheckIfThereIsACorrigedumIThinkThereMayBe:upward}.

Of course, the litmus test of NP refinements such as UP, $\up_{\leq k}$,
and FewP is {\em the extent to which they allow us to refine
the upper bounds on the complexity of natural NP problems}.
Of these classes, UP has been most successful in this regard.
UP is known to provide an upper bound on the complexity of (a 
language version of) the discrete logarithm 
problem~\cite{gro-sel:j:complexity-measures}, and UP (indeed
$\up \inter \coup$) is known to provide an upper bound on the 
complexity of 
primality testing~\cite{fel-kob:c:self-witnessing}.

However, there are certain NP problems whose richness of 
structure has to date defied attempts to put them in UP or even FewP,
yet that nonetheless intuitively seem to use less than the full 
generality of NP's acceptance mechanism.  To try to categorize these 
problems, we introduce the class EP, which is intermediate 
between FewP and~NP: \mbox{$\fewp \seq \ep \seq \np$}.  In particular,
EP is the NP subclass whose acceptance set is
$\{2^i \condition i \in \naturalnumber\}$, $\naturalnumber =
\{0,1,2,3,\ldots\}$.  

In Section~\ref{s:concrete}, we 
provide improved
upper bounds on the complexity of the problems {OBDD}
(Ordered Binary Decision Diagram)
{Negation} {Equivalence},
{2-Dag} {Interchange} {Equivalence}, and 
{Boolean} {Negation} {Equivalence}.
These three problems are trivially in, respectively,
NP, NP, and $\np^\np$.
We provide, respectively, EP, EP, and $\ep^\np$ 
upper bounds.
The problems are not known to
belong to (and do not seem to obviously belong to),
respectively, FewP, FewP, and $\fewp^\np$.   

In Section~\ref{s:fewp}, we prove a general result regarding 
containment of FewP in certain restricted counting classes.
In particular, we establish a sufficient condition for when
restricted counting classes 
contain~FewP\@.
From our result it follows that 
EP contains FewP and,
moreover,
our result subsumes as special cases some previously known results
from the literature.

In Section~\ref{s:open-q}, we list some open questions related to our
work.

\section{Concrete Problems and EP}\label{s:concrete}
In this section, we provide concrete problems known to be in NP 
(or $\npnp$), and we prove they are in fact in EP (or $\ep^\np$).
We now define the class EP (mnemonic: the number of accepting 
computation paths is restricted to being either 0 or some power
(some {\em{}e}xponentiation) of~2).
For any nondeterministic polynomial-time Turing machine $N$ and any
string~$x$, let $\mbox{\#acc}_{N}(x)$ denote the number of accepting 
computation paths of $N$ on input~$x$. 
Our alphabet $\Sigma$ will be $\{0,1\}$. For any string~$x \in \sigmastar$,
let $|x|$ denote the length of~$x$. 
\begin{definition}
  $\ep$\/~denotes the class of all languages $L$ for which there is a
  nondeterministic polynomial-time Turing machine
  $N$ such that, for each input $x \in \sigmastar$,
\begin{eqnarray*}
x \not\in L & \implies & \mbox{\#acc}_{N}(x) = 0, \mbox{ and} \\
x \in L     & \implies & \mbox{\#acc}_{N}(x) \in \{ 2^i
\condition i \in \naturalnumber \}.
\end{eqnarray*}

\end{definition}

Consider the following well-known problem.

\medskip

{\samepage

\noindent {\bf Problem:}~~{Boolean} 
{Negation} {Equivalence} ({BNE}) 
(see the survey by Harrison~\cite{har:b:boolean-survey} and
the bibliography provided after the references in the paper by
Borchert, Ranjan, and
Stephan~\cite{bor-ran-ste:j:boolean-equivalence})
\\
\noindent {\bf Input:}~~Two boolean functions
(input as boolean formulas using variable names 
and the symbols $\{\wedge,\vee,\neg,(,)\}$), 
$f(x_1,\ldots,x_n)$ and 
$g(x_1,\ldots,x_n)$, over the same $n$ boolean 
variables.%
\\
\noindent {\bf Question:}~~Are $f$ and $g$ negation equivalent?  That is,
can one negate some of the inputs of
$g$ such that $f$ and the modified function $g'$ are 
equivalent?\footnote{The notion of boolean function
equivalence underlying the definition of negation equivalence is the
standard one.   Two boolean functions (over the same variables) are 
equivalent if they have the same truth value for every 
assignment to their variables.   Testing equivalence of pairs of 
boolean formulas is in coNP.}

For concreteness as a language problem, $\bne = 
\{ (f,g) \condition f$ and $g$ are negation equivalent$\}$.

} %

\medskip

For example, the two boolean functions described by the formulas
$x_1\lor x_2\lor x_3$ and $x_1\lor \neg x_2\lor \neg x_3$ are negation
equivalent by negating $x_2$ and~$x_3$.  Regarding lower bounds,
Borchert, Ranjan, and Stephan~\cite{bor-ran-ste:j:boolean-equivalence}
have shown that $\bne$ is US-hard~\cite{bla-gur:j:unique-sat}, and
thus in particular is coNP-hard.  Regarding upper bounds, $\bne \in
\npnp$~\cite{bor-ran-ste:j:boolean-equivalence} and $\bne\in\coam^\np$
(combining~\cite{bor-ran-ste:j:boolean-equivalence}
and~\cite{agr-thi:c:boolean-isomorphism}).  It follows from the latter
that $\bne$ is not $\npnp$-complete unless the polynomial hierarchy
collapses~(\cite{agr-thi:c:boolean-isomorphism}, in light
of~\cite{bor-ran-ste:j:boolean-equivalence,sch:j:pr-low}).
Interestingly, neither of these two upper bounds---$\npnp$
and $\coam^\np$---is known to imply the other.

We now prove
$\bne \in \ep^\np$, which is neither known to imply nor known to
be implied by the
$\coam^\np$ upper bound, but which clearly 
improves the $\npnp$ upper bound as
$\ep^\np \seq \npnp$.
\begin{theorem}\label{t:xz}
$\bne \in\ep^\np$. 
\end{theorem}

\noindent {\bf Proof}.  \quad
Suppose a given instance of $\bne$ consists of $f$ and $g$, 
each over the variables $x_1,\ldots, x_n$.  
A negation of some of the input variables of $g$ as in the
definition of $\bne$ can be represented by a vector $\vec{v}=(c_1,\dots,c_n)$
in the vector space ${\rm GF(2)}^n$,
where each $c_i$ is either $0$ or~$1$ and $c_i=1$ means that the
variable $x_i$ will be negated. Let $g_{\vec{v}}$ be the boolean function
resulting from $g$ after the application of the negations described by
$\vec{v}$, i.e., $g_{\vec{v}}(\vec{u}) = g(\vec{v} + \vec{u})$.  
Now it is easy to see (double negation equals identity,
and addition in ${\rm GF(2)}^n$ is associative) that,
for each fixed boolean function~$g$, the set of negation vectors $\vec{v}$
such that $g$ equals $g_{\vec{v}}$ is a 
linear subspace $V_g$ of 
${\rm GF(2)}^n$.
It is not hard to see that if $\vec{w}$ is {\em any\/} negation
vector such that $f=g_{\vec{w}}$, then the affine subspace $\vec{w}+V_g$ is
the set of {\em all\/} 
negation vectors witnessing the negation equivalence of
$f$ and~$g$.   
Of course, $\vec{w}+V_g$ will be of the same cardinality as the 
subspace $V_g$
(as addition by $\vec{w}$ induces a bijection between ${\rm GF}(2)^n$
and itself),
and as an $\ell$-dimensional vector space over the field ${\rm GF(2)}$
has exactly $2^{\ell}$ vectors, 
$\vec{w}+V_g$ will contain exactly
$2^m$ vectors, where $m$ is the dimension of~$V_g$.  
So the following nondeterministic program shows that {BNE} is in EP with
an NP oracle: Read the two input functions $f$ and $g$ (checking that
they are both over the same number of variables and that the variables
have the same naming scheme), guess a negation vector $\vec{v}$ and
accept if and only if the oracle confirms that $f$ is
equal to $g$ altered by
the negation vector~$\vec{v}$.  This shows that {BNE} is
in $\ep^\np$, since if $f$ and $g$ are not negation equivalent, then
there is no accepting path, and otherwise there are exactly $2^m$
accepting paths, where $m$ is the dimension of the affine subspace
discussed above.~\qed

\medskip

There are ways of describing boolean functions such that the
equivalence problem is in~P\@.  The most prominent 
such way is by ordered
binary decision diagrams (OBDDs).\footnote{Fortune, Hopcroft,
and Schmidt~\protect\cite{for-hop-sch:c:equivalence} 
were the ones who proved that 
equivalence for OBDDs is in $\p$.  
\mbox{OBDDs} have recently become a structure of interest to 
theoretical computer scientists in 
a variety of settings, see, 
e.g.,~\protect\cite{tak-nou-yaj:c:obdd-threshold-function}.
Recently, Feigenbaum et
al.~\protect\cite{fei-kan-var-vis:c:graphs-represented-by-OBDDs} 
showed that
when one 
modifies
certain graph problems, such as Independent Set or
Graph Accessibility,
so that the input graph is succinctly
represented as an OBDD,
there is an exponential blow-up
in 
complexity.
For general
background on \mbox{OBDDs} see, for example,
the survey by Bryant~\protect\cite{bry:j:obdd-survey}.}
So, essentially by the same type of discussion found in 
the proof of Theorem~\ref{t:xz}, the following computational problem,
{OBDD} {Negation} {Equivalence}, is in (nonrelativized)~EP:
Given a pair $(e,f)$ of OBDDs, are the boolean functions described by
$e$ and $f$ negation equivalent?

If we consider the special case that for the two OBDDs $(e,f)$ above the
order of the variables is required to be the same, we see that the
following graph-theoretic problem is in (nonrelativized)~EP\@.  A
{\em 2-dag\/} is a directed acyclic graph (without labels) with a
unique root and either 0 or 2 ordered successors for each node. For a
2-dag each node is assigned a depth, namely the distance to the root.
Now consider the following computational problem ({2-Dag} 
{Interchange} {Equivalence}): Given two 2-dags $F$
and $G$, is there a sequence of natural numbers $(i_1,\dots, i_m)$ such
that, if in $G$ for each node of depth $i_1,\dots, i_m$ its two
successors (if they exist) are interchanged, then the modified 2-dag
$G'$ equals~$F$?  This problem can be shown to be in EP (similarly
to the argument above).  Moreover, the problem can easily be reduced
to {Graph} {Isomorphism}.  The authors know of no
P algorithm for the general case of {2-Dag} {Interchange} 
{Equivalence}, 
though 
the special case of
this problem with binary trees instead of general 2-dags has an easy 
deterministic polynomial-time algorithm.

\section{Location of EP}\label{s:fewp}

\subsection{Result}
We state a general result that our technique gives, regarding
the containment of FewP in restricted counting classes.  We 
need some additional definitions.  
\begin{definition}
Let $S$ be any set of positive integers.
Define the restricted counting class $\rc_S$ 
as follows.
$L \in \rc_S$ if and only if there exists a nondeterministic polynomial-time
Turing machine $N$ such that, for every $x \in \sigmastar$, 
\begin{enumerate}
\item if $x \in L$ then $\mbox{\#acc}_N(x) \in S$, and
\item if $x \not\in L$ then $\mbox{\#acc}_N(x) = 0$.
\end{enumerate}
\end{definition}

For example, Valiant's 
extensively studied class UP equals $\rc_{\{1\}}$, 
and, for each $k\geq 2$,
the class ${\rm ModZ}_k\p$ 
of Beigel, Gill, and 
Hertrampf~\protect\cite{bei-gil-her:cOUTbybei-gil-jour-and-her-jour:mod} equals
$\rc_{\naturalnumber - 
\{ a \mid (\exists b \in \naturalnumber)\, [a = b\cdot k] \}}$.

A set is non-gappy if it has only small holes.

\begin{definition}
\label{d:non-gappy}
Let $S$ be any set of positive integers.
We say $S$ is {\em non-gappy\/} if 
$S \neq \emptyset$ and $
(\exists k > 0)(\forall n \in S)(\exists m \in S)[ m > n \land
m/n \leq k]$.
\end{definition}

\begin{definition} {\rm{}\cite{har-yes:j:computation}} \quad
Let $L$ be any subset of $\sigmastar$.
We say $L$ is {\em $\p$-printable\/} if 
there is a deterministic Turing machine $M$ that runs in polynomial-time
such that, for every nonnegative integer $n$, $M(0^n)$ prints 
out the set $\{ x \condition x \in L \land |x| \leq n\}$.
\end{definition}

\begin{theorem}\label{t:oe}
  Let $T$ be any set of positive integers such that $T$ has a 
  non-gappy, $\p$-printable subset.
  Then $\fewp \subseteq 
  \rc_T$.\footnote{Though this result 
is stated in a relatively
general format, we mention in 
passing that even the restriction employed can be
relaxed to the case of nonempty 
sets of positive integers for which, for some uniform constant,
given
any integer in the set finding another larger but at 
most multiplicatively-constantly-larger
integer in the set is a polynomial-time task.  One can even 
slightly relax the growth rate, but one has to be very careful to
avoid a ``bootstrapping'' growth-explosion effect via 
clocking growth rates always with respect to the input.  In any case,
we feel the current statement of the theorem is general enough to 
capture the generality of the result without being so technical as to 
obscure its essence.}
\end{theorem}

Our proof technique builds (e.g., by adding a rate-of-growth
argument) on that used by Cai
and Hemachandra~\cite{cai-hem:j:parity}
to prove $\fewp\seq\parityp$, where 
$\parityp$~\cite{gol-par:j:ep,pap-zac:c:two-remarks} is the class
of languages $L$ such that for some nondeterministic polynomial-time
Turing machine~$N$, 
on each $x$ it holds that $x\in L \iff \#acc_N(x) \equiv 1~({\rm mod}~2)$.  
We note that K\"obler, Sch\"oning, Toda,
and Tor\'{a}n~\cite{koe-sch-tod-tor:j:few} interestingly built 
on that technique in their proof that~$\fewp\seq\ceqp$, where
$\ceqp$~\cite{wag:j:succinct}
is the class of languages $L$ such that there is a polynomial-time
function $f$ and a nondeterministic polynomial-time 
Turing machine $N$ such that for each~$x$, $x\in L$ if and 
only if $\#acc_N(x) = f(x)$.

\medskip

\noindent
{\bf Proof of Theorem~\ref{t:oe}}.
\quad
\newcommand{\binomFOO}[2]{{{ {{#1}} \choose {{#2}}   }}}
Let $S$ be a non-gappy, P-printable subset of~$T$. Let $k > 0$ be, 
for $S$, some constant satisfying 
Definition~\ref{d:non-gappy}. 

Let $L$ be any language in $\fewp$. Let $\hat{N}$ be a machine witnessing
that $L \in \fewp$, and let $p$ be a polynomial bounding the
nondeterministic ambiguity of~$\hat{N}$, i.e., 
for each input~$x$,
$\mbox{\#acc}_{\hat{N}}(x) \leq
p(|x|)$.
To show that $L \in \rc_T$, we describe
a nondeterministic polynomial-time Turing machine $N$ that accepts
$L$ via the $\rc_T$ acceptance mechanism.

On input~$x$, $N$ chooses $p(|x|)$ natural numbers $c_1, c_2, \ldots ,
c_{p(|x|)}$ as follows. Initially, we assume that~$c_1$, which is
defined to be the least element of~$S$, is hard-coded into the program
of~$N$. Successively, for $i = 2, \ldots , p(|x|)$, machine $N$ on input
$x$ does the following:
\begin{itemize}
\item Let $c_1, \ldots , c_{i-1}$ be the constants that have already
  been chosen.  Define 
\[
b_i = \binomFOO{i}{1} c_1 + \binomFOO{i}{2} c_2 + \cdots + \binomFOO{i}{i-1}
  c_{i-1}.
\]

\item Let $a_i$ be the least element of $S$ such that $b_i \leq a_i$.

\item Set $c_i = a_i - b_i$.
\end{itemize}
After having chosen these constants, $N$ (still on input $x$) will do the
following: 
Nondeterministically guess an integer $i \in \{1,2, \ldots , p(|x|)\}$ and, 
for
each $i$ guessed, nondeterministically guess each 
(unordered) $i$-tuple of
distinct paths of~$\hat{N}(x)$.  On each path~$\alpha$ resulting from such a
guess series, $N(x)$ sees whether the $i$ paths of $\hat{N}(x)$ that were
guessed on $\alpha$ are all accepting paths.  If all are accepting
paths, then path~$\alpha$, via trivial 
nondeterministic guesses, splits itself into $c_i$
accepting paths.  On the other hand, if at least one of the $i$
guessed paths is a rejecting path, then path $\alpha$ simply rejects.
This completes the description of~$N$.

The
intuition behind the construction of $N$ is that for each input $x$
the following holds.  $N(x)$ has $c_1$ accepting paths for each
accepting path of~$\hat{N}(x)$; $N(x)$ has $c_2$ additional accepting paths
for each pair of distinct accepting paths of~$\hat{N}(x)$; and so on.
So, if $x\in L$, 
$N(x)$ has $c_{\mbox{\#acc}_{\hat{N}}(x)}$ additional accepting
paths for the (one) 
$\mbox{\#acc}_{\hat{N}}(x)$-tuple of distinct accepting paths
of~$\hat{N}(x)$. However, if for some $z$ with $\mbox{\#acc}_{\hat{N}}(x) < z \leq
p(|x|)$ a $z$-tuple
of distinct paths of $\hat{N}(x)$ was guessed on a path $\alpha$ of~$N(x)$,
then $\alpha$ must contain a rejecting path of~$\hat{N}(x)$, and thus $N(x)$
will have no accepting paths related
to $c_z$. This intuition is expressed
formally by:
\[
\mbox{\#acc}_{N}(x) = 
{\binomFOO{\mbox{\#acc}_{\hat{N}}(x)}{1}}  c_1 + 
{\binomFOO{\mbox{\#acc}_{\hat{N}}(x)}{2}}  c_2 + \cdots + 
{\binomFOO{\mbox{\#acc}_{\hat{N}}(x)}{\mbox{\#acc}_{\hat{N}}(x)}} 
c_{\mbox{\#acc}_{\hat{N}}(x)}.
\]

Assume $x \in L$. Thus, $0 < \mbox{\#acc}_{\hat{N}}(x) \leq p(|x|)$. Since
$c_{\mbox{\#acc}_{\hat{N}}(x)}$ was chosen such that
\begin{eqnarray*}
\mbox{\#acc}_{\hat{N}}(x) = 1 & \implies &
\mbox{\#acc}_{N}(x) = c_1 , \mbox{ and}\\
\mbox{\#acc}_{\hat{N}}(x) \geq 2 & \implies &
\mbox{\#acc}_{N}(x) = 
b_{\mbox{\#acc}_{\hat{N}}(x)} + c_{\mbox{\#acc}_{\hat{N}}(x)} = a_{\mbox{\#acc}_{\hat{N}}(x)} ,
\end{eqnarray*}
and since both $c_1$ and $a_{\mbox{\#acc}_{\hat{N}}(x)}$ are elements
of~$S$, it follows that $\mbox{\#acc}_{N}(x) \in T$. On the other
hand, if $x \not\in L$ then $\mbox{\#acc}_{\hat{N}}(x) = 0$, and so
$\mbox{\#acc}_{N}(x) = 0$.  

So now, to prove that $L \in \rc_T$, it suffices to establish an
exponential (in~$|x|$) upper bound on the value of 
$\max_{i \leq p(|x|)} c_i$.

We will consider, for $j \geq 2$, what bounds hold on 
the value of $c_j$.
By construction of $N$ and
since $S$ is non-gappy, we have $c_j \leq a_j \leq k  b_j$.
Regarding the latter inequality, note that $b_j$ is not necessarily an
element of~$S$.  However, for each~$j$, $c_1 \leq b_j$; so for
each~$j$, there exists a $\hat{b}_j \in S$ such that $\hat{b}_j \leq b_j$ and
$\hat{b}_j$ is the greatest such integer in~$S$. Since $a_j$ is defined to
be the least element of $S$ such that $b_j \leq a_j$, we have $a_j
\leq k  \hat{b}_j \leq k  b_j$.

From the above and the definition of~$b_j$, we have:
\begin{eqnarray}
\label{equ:estimate}
\nonumber
c_j & \leq & k 
  \left( \binomFOO{j}{1} c_1 + \binomFOO{j}{2} c_2 + \cdots +
  \binomFOO{j}{j-1} c_{j-1} \right) \\ 
    & \leq & k (j-1) \binomFOO{j}{\lceil \frac{j}{2} \rceil} 
             \max_{1 \leq i \leq j-1} c_i .
\end{eqnarray}
The factor $j - 1$ in inequality~(\ref{equ:estimate}) is the number of
terms in~$b_j$, and the coefficient $\binomFOO{j}{\lceil \frac{j}{2}
\rceil}$ is the biggest binomial coefficient of any term in~$b_j$.

Recall that once we were given $S\seq T$ we fixed $k$.
For all sufficiently large~$j$ the following holds:
\begin{equation}
\label{equ:ungl}
k  (j-1) \binomFOO{j}{\lceil \frac{j}{2} \rceil} \leq
{\binomFOO{j}{\lceil \frac{j}{2} \rceil}}^2 \leq 
{\left(2^j\right)}^2.
\end{equation}
In particular, let $j_{bad} = j_{bad}(k)$ be the largest $j$ for 
which the above inequality fails to hold (if it always holds, 
set $j_{bad}=1$).  Let $I_{bad} =
\max_{1\leq i \leq j_{bad}} c_i$.
From inequalities~(\ref{equ:estimate}) and~(\ref{equ:ungl}),
we clearly have that,
for $j>j_{bad}$:
\[
c_j \leq  I_{bad} \cdot \prod_{j_{bad} < i \leq j} 2^{2i},
\]
and, 
for $j \leq j_{bad}$,
$c_j \leq I_{bad}$.
This implies that 
\mbox{$c_j = 2^{{\cal O}(j^2)}$}.  

Thus, for the fixed $k$ associated with $S\seq T$, the value of 
$\max_{i \leq p(|x|)} c_i$ indeed is bounded by an exponential
function in $|x|$.
Hence, $L \in \rc_T$, and thus $\fewp \seq \rc_T$.~\qed

\medskip

It is immediate from its definition that $\ep\seq\np$.  It is 
also clear that the quantum-computation-related class
$\ceqphalf$ of 
Berthiaume and Brassard~\cite{ber-bra:c:quantum-survey}
is contained in EP\@.\footnote{%
$\ceqphalf$, introduced by Berthiaume and 
Brassard~\protect\cite{ber-bra:c:quantum-survey} in their 
study of quantum complexity, is a variant of 
the class WPP of
Fenner, Fortnow, and Kurtz~\protect\cite{fen-for-kur:j:gap}.  
Namely, $\ceqphalf$
is the class of languages $L$ such that there is some nondeterministic
Turing machine such that if the input is in $L$ exactly half of 
the paths are accepting paths and if the input is not in $L$ 
none of the paths are accepting paths.
\label{foo:ceqphalf}
}
From Theorem~\ref{t:oe} it
immediately follows
that $\fewp \subseteq \ep$,
since $\ep = \rc_{\{ 2^i \mid i \in \naturalnumber \}}$ and
$\{2^i \condition i \in \naturalnumber \}$ is clearly a P-printable,
non-gappy set.

\begin{corollary}
\label{c:fe}
$\fewp \seq \ep$.
\end{corollary}

The comments attached to our on-line technical report
version~\cite{bor-hem-rot:tECCCoutdatedExceptOkToCiteForHistoryOfComments}
give some of the history of the proof of our results and of some 
valuable comments made by Richard Beigel, in particular that 
FewP is also contained in the 
EP analog based on any integer~$n$ (note that the acceptance
sets for such classes  
are P-printable and non-gappy).

Cai and Hemachandra's result
$\fewp\seq\parityp$~\cite{cai-hem:j:parity} has been generalized to
$\fewp \seq {\rm ModZ}_k\p$, for each 
$k \geq 2$~\cite{bei-gil-her:cOUTbybei-gil-jour-and-her-jour:mod}. 
This generalization
also follows as a special case of Theorem~\ref{t:oe}, since ${\rm
ModZ}_k\p = \rc_{\naturalnumber - \{ a \mid (\exists b \in
\naturalnumber )\, [a = b\cdot k] \}}$ as mentioned above. 

\begin{corollary}
{\rm{}\cite{bei-gil-her:cOUTbybei-gil-jour-and-her-jour:mod}}\quad
For each $k \geq 2$, $\fewp \seq {\rm ModZ}_k\p$.
\end{corollary}

\subsection{Discussion}

An immediate question is how Corollary~\ref{c:fe} 
relates to known results about $\fewp$.  Clearly,
Corollary~\ref{c:fe} represents an improvement on the trivial 
inclusion $\fewp\seq \np$.  However, how does it compare with 
the nontrivial result of K\"obler et al.~\cite{koe-sch-tod-tor:j:few}
and Fenner, Fortnow, and Kurtz~\cite{fen-for-kur:j:gap}
that $\fewp\seq\few\seq\spp\seq\parityp\inter
\ceqp$?
Informally stated, Few~\cite{cai-hem:j:parity} 
is what a P machine can do given one call
to a $\sharpp$ function that obeys the promise that its value is
always at most polynomial.  
SPP~\cite{fen-for-kur:j:gap,hem-ogi:j:closure}
is the class of sets $L$ such that for some nondeterministic
polynomial-time Turing machine $N$ it holds that 
if $x\not\in L$ then $N(x)$ has one fewer accepting path than it has rejecting
paths, and if $x\in L$ then the numbers of accepting and rejecting
paths of $N(x)$ are equal. Curiously, note that the nontrivial
result that
$\fewp\seq\spp$ itself 
neither is known to imply nor is known to be implied by the
trivial result $\fewp\subseteq \np$.

There are a number of related aspects to the question raised above.  
First, is $\spp\subseteq \ep$? This inclusion---which would make 
Corollary~\ref{c:fe}
a trivial consequence of the known result $\fewp\seq\spp$---seems 
unlikely, as if $\spp\seq\ep$, then $\spp\seq \np$, 
and $\spp\seq\np$ is considered unlikely 
(see~\cite{fen-for-kur:j:gap,ogi-tod:j:counting-hard}).
Second, is $\ep\seq\spp$? (This inclusion would make 
Corollary~\ref{c:fe} a strengthening of the known result
that $\fewp\seq\spp$.)~~We do not know.  Third,
notice that we proved $\fewp\seq\ep$ but 
that the 
K\"obler et al.~\cite{koe-sch-tod-tor:j:few}
and Fenner, Fortnow, and Kurtz~\cite{fen-for-kur:j:gap}
work shows that $\few\seq\spp$.  Can our result be
extended to show $\few\seq\ep$?  
The reason we mention this is that 
often it is the case that when one can prove something about FewP,
then one can also prove it about the slightly bigger class Few.
For
  example, Cai and Hemachandra, after showing that FewP is in
  $\parityp$, then easily applied their technique to show that even Few
  is in~$\parityp$~\cite{cai-hem:j:parity}.  
Similarly, it is immediately
  clear that FewP has Turing-complete sets if and only if
  Few has Turing-complete sets, and so the proof that 
  there is a relativized world in which FewP lacks Turing-complete
  sets~\cite{hem-jai-ver:j:up-turing} implicitly proves
  that there is a world in which Few lacks Turing-complete
  sets (see also~\cite{ver:j:oracle-survey}).
  However, in the
  case of Corollary~\ref{c:fe}, it is unlikely that by 
  modifying the technique in a way similar to that done
  by Cai and Hemachandra one could hope to establish the slightly
  stronger result that EP even contains Few.  Why?
  Clearly $\coup
  \seq \few$ and $\ep \seq \np$, so the assumption $\few \seq \ep$ would
  imply (along with other even more 
  unlikely things) %
  $\coup \seq \np$.

Fourth, one might wonder directly, since $\fewp\seq\parityp$ is 
known, about the relationship between $\ep$ and $\parityp$.
That is, how is $\ep$ (powers-of-two acceptance) related
to $\parityp$ (multiples-of-two 
acceptance).\footnote{%
\label{foot:promise}%
However,
one should keep in mind the contrasting rejection sets of these 
two classes.}
We note the following.  By a diagonalization so routine as to 
not be worth including here,
one can show $(\exists A)\,[\coup^A\not\seq\ep^A]$.
It follows immediately, since (for each $B$) $\coup^B 
\subseteq \few^B$,  that 
$(\exists A)\,[\few^A\not\seq\ep^A]$ and
$(\exists A)\,[\parityp^A\not\seq\ep^A]$.
Similarly, if one looks at the test language inside the 
proof of Proposition~12 of Beigel's 1991 ``mod classes''
paper~\cite{bei:j:mod}, one can see that for his case ``$k=2$''
the test language is in 
(relativized) $\ceqphalf$,
and thus as a corollary
to his proof one can claim
$(\exists A)\,[\ceqphalf^A\not\seq\parityp^A]$.
It follows immediately that 
$(\exists A)\,[\ep^A\not\seq\parityp^A]$.
Since these are standard diagonalizations that can easily be 
interleaved, it is easy to see that there is 
a relativized world in which EP and $\parityp$ are 
incomparable (i.e., neither is contained in the 
other).\footnote{%
Concerning ${\rm Mod}_q\p$ classes for values
$q>2$~\protect\cite{cai-hem:j:parity,bei-gil-her:cOUTbybei-gil-jour-and-her-jour:mod},
it follows easily
from the known relations among such 
classes~\protect\cite{bei-gil-her:cOUTbybei-gil-jour-and-her-jour:mod}
and the obvious fact that 
powers of 2 are never
congruent to zero 
modulo $j$, where $j$ is any number greater than $2$
that is not a power of two, that EP is contained in ${\rm Mod}_q\p$
for all $q>2$ such that $q$ is not a power of two.  ($q$ values
that are powers of two give just $\parityp$ and thus as noted
above are incomparable to EP in some relativized world.)  So it 
also follows from this, in light of 
Beigel's~\protect\cite{bei:j:mod} result that
for any distinct primes $q_1$ and $q_2$
there are oracles relative to
which
${\rm Mod}_{q_1}\p$ and 
${\rm Mod}_{q_2}\p$ are 
incomparable (i.e., neither contains the other), 
that for every $q>1$ there is
an oracle such that 
${\rm Mod}_{q}\p$ is not contained in EP.
}%

Fifth and finally, 
to complete this discussion, what is the relation between EP
and~$\ceqp$? 
Proposition~\ref{p:es} below shows that EP is contained
in~$\ceqp$.\footnote{After seeing an earlier draft of this paper, 
Richard Beigel has communicated (February, 1998) to the authors 
that he observed that EP is even contained in the class
LWPP~\protect\cite{fen-for-kur:j:gap}. Since it is known from the 
work of Fenner, Fortnow, 
and Kurtz~\cite{fen-for-kur:j:gap} that $\spp \seq \mbox{\rm LWPP} \seq
\ceqp$, this improves upon our result and in particular shows that
EP is PP-low (i.e., $\pp = \pp^{\rm EP}$), where PP denotes
probabilistic polynomial time.
}
Thus,
Corollary~\ref{c:fe} improves upon
K\"obler et al.'s result that
$\fewp\seq\ceqp$~\cite{koe-sch-tod-tor:j:few}---an improvement that
seems to neither imply nor be implied by
other improvements of their result such
as $\few \seq \spp$~(\cite{koe-sch-tod-tor:j:few}, see
also~\cite{fen-for-kur:j:gap}).

\begin{proposition}
\label{p:es}
$\ep \seq \ceqp$.
\end{proposition}

\noindent
{\bf Proof}.
\quad
Let $\es$ (which is the nonpromise
version of EP) denote the 
class of all languages $L$ for which there is a
nondeterministic polynomial-time Turing machine $N$ such that, for
each input $x \in \sigmastar$, $x \in L \iff \mbox{\#acc}_{N}(x) \in
\{ 2^i \condition i \in \naturalnumber \}$. Note that, clearly,
$\ep \subseteq \es$.  However, note that $\es = \ceqp$ 
as we now argue.
 $\es \seq \{ L \condition (\exists A \in \ceqp)[L \leq_d^p 
A]\}$ is immediately
 clear from the definitions, where $\leq_d^p$ is polynomial-time
 disjunctive reducibility.
 So $\es\seq\ceqp$, as it is known that 
 $\ceqp = \{ L \condition (\exists A \in \ceqp)[L \leq_d^p A]\}$
 \protect\cite{bei-cha-ogi:j:difference-hierarchies}.
 To show $\ceqp\seq\es$, consider a $\ceqp$ machine
 and the function $f$ giving the number of paths on
 which it would accept.  Let $w(x)$ be the smallest integer
 such that $2^{w(x)} > f(x)$.  Consider the EP machine that on
 input $x$ has $2^{1+w(x)} - f(x)$ paths that immediately accept, and 
 that also has paths that simulate the $\ceqp$ machine.  Note that
 this machine accepts the $\ceqp$ language.~\qed

\section{Open Questions}
\label{s:open-q}
Does $\ep$ equal $\np$?  It would be nice to give evidence that such
an equality would, for example, collapse the polynomial hierarchy.
However, $\up\subseteq\ep\seq\np$, and at the present time, it is open
whether even the stronger assumption $\up=\np$ implies any startling
collapses.  
Also, does EP, in contrast to most promise classes, have complete sets?
We conjecture that EP lacks complete sets (of course, if EP equals NP
then EP has complete sets).

It is clear that EP is closed under conjunctive reductions and under
disjoint union, and (thus) under intersection.  Is EP closed under
disjunctive reductions or union?

Finally, define:
\begin{definition}\label{d:forth}
$\listp_{\cal F}$ is the class of all sets $L$ such that
\mbox{$(\exists f\in\fp,~f: \sigmastar \rightarrow 2^{\naturalnumber})$}
\mbox{$(\exists h\in {\cal F})$} 
$( \exists$ nondeterministic polynomial-time
Turing machine $N) \, (\forall x)$ 
$$[
||f(x)|| \leq h(|x|) \land \mbox{\#acc}_N(x)\in f(x) \union \{0\}
\land \left( x\in L \iff \mbox{\#acc}_N(x)\geq 1\right)],$$ 
where
$\fp$ denotes the class of (total) polynomial-time functions.
\end{definition}
Let $\listp = \listp_\poly$, where $\poly$ is the set of all 
polynomials.  That is, ListP is very similar to EP, except the list 
of potential numbers of accepting paths on an input, rather
than being $\{0,1,2,4,8,\ldots\}$, is instead some polynomial-time
computable polynomial-sized list (of numbers written in binary)
that may depend on the input.  Clearly, $\ep\seq\listp$, and in fact
the EP analogs based not on powers of 2 but on powers of $k$ are 
also in ListP\@.  Is \mbox{$\sat \in \listp$}?   We do not know.  
However, note that ListP is in some sense a language analog of the 
function-based
notion of ``enumerative counting''~\cite{cai-hem:j:enum}.  
It follows 
immediately from an enumerative counting result of Cai and 
Hemachandra~\cite{cai-hem:j:approx2} and, independently,
Amir, Beigel, Gasarch, and 
Toda (included in~\cite{ami-bei-gas:c:uni})
that SAT is in ListP {via a machine $M$ that uses
the canonical witness scheme for $\sat$ (or any 
witness scheme whose numbers of witnesses are 1-Turing
interreducible with those of the canonical witness scheme
in the context of the input)\/} if
and only if $\p = \psharpp$.  So if we knew that 
all witness schemes for SAT were closely related to 
the canonical one, then we would know that SAT was in ListP
if and only if $\p = \psharpp$.  However, can SAT 
thwart this by having some
bizarre witness scheme deeply unrelated to its canonical
witness scheme?  In fact, Fischer, Hemaspaandra, 
and Torenvliet have recently provided 
sufficient conditions for such schemes to 
exist~\cite{fis-hem-tor:b:witness-iso}.

Similar comments apply to {Graph} {Automorphism} ($\ga$).
Is $\ga \in \listp$?   We do not know.  
However, it follows 
immediately from a result of 
Chang, Gasarch, and 
Tor{\'{a}}n~\cite{cha-gas-tor:c:graph-auto}
that {GA} is in ListP {via a machine $M$ that uses
the canonical witness scheme for {GA} (or any 
witness scheme whose numbers of witnesses are 1-Turing
interreducible with those of the canonical witness scheme
in the context of the input)} only if
{Graph} 
{Isomorphism}~[sic] is in R,
random 
polynomial time.

{\samepage
\begin{center}
{\bf Acknowledgments}
\end{center}
\nopagebreak
\noindent 
We thank Richard Beigel, Frank Stephan, 
and Gerd Wechsung for interesting comments or discussions,  and
Dieter Kratsch, Haiko M\"{u}ller, and 
Johannes Waldmann for very kindly 
letting us use their office's computers to type in 
part of this paper.
}%

{\small

{\small  

\bibliography{gry}

}

}

\end{document}